\documentclass{emulateapj}
\usepackage{apjfonts}

\newcommand{\Msun}      {\mbox{$\rm\,M_{\mathord\odot}$}}

\submitted{Accepted by the Astrophysical Journal Letters}

\begin{document}

\lefthead{Quiescent Black Holes}
\righthead{J.A. Tomsick et al.}

\title{{\em Chandra} Detections of Two Quiescent Black Hole X-Ray Transients}

\author{John A. Tomsick\altaffilmark{1},
St\'ephane Corbel\altaffilmark{2}, 
Rob Fender\altaffilmark{3}, 
Jon M. Miller\altaffilmark{4,5}, 
Jerome A. Orosz\altaffilmark{6}, 
Michael P. Rupen\altaffilmark{7},
Tasso Tzioumis\altaffilmark{8},
Rudy Wijnands\altaffilmark{9},
Philip Kaaret\altaffilmark{4}}

\altaffiltext{1}{Center for Astrophysics and Space Sciences, Code
0424, University of California at San Diego, La Jolla, CA,
92093, USA (e-mail: jtomsick@ucsd.edu)}

\altaffiltext{2}{Universit\'e Paris VII and Service d'Astrophysique, 
CEA Saclay, 91191 Gif sur Yvette, France}

\altaffiltext{3}{Astronomical Institute ``Anton Pannekoek,'' University
of Amsterdam and Center for High Energy Astrophysics, Kruislaan 403, 
NL-1098 SJ Amsterdam, The Netherlands}

\altaffiltext{4}{Harvard-Smithsonian Center for Astrophysics, 60 Garden Street,
 Cambridge, MA, 02138, USA}

\altaffiltext{5}{National Science Foundation Astronomy and Astrophysics Fellow}

\altaffiltext{6}{Department of Astronomy, San Diego State University, San 
Diego, CA, 92182, USA}

\altaffiltext{7}{National Radio Astronomy Observatory, Soccoro, NM 87801}

\altaffiltext{8}{ATNF, CSIRO, P.O. Box 76, Epping, NSW 1710, Australia}

\altaffiltext{9}{School of Physics and Astronomy, University of St. 
Andrews, North Haugh, St. Andrews KY16 9SS, UK}

\begin{abstract}

Using the {\em Chandra X-ray Observatory}, we have detected
the black hole transients V4641~Sgr and XTE J1859+226 in
their low luminosity, quiescent states.  The 0.3-8 keV
luminosities are $(4.0^{+3.3}_{-2.4})\times 10^{31}$ 
($d$/7 kpc)$^{2}$ erg~s$^{-1}$ and $(4.2^{+4.8}_{-2.2})\times 
10^{31}$ ($d$/11 kpc)$^{2}$ erg~s$^{-1}$ for V4641~Sgr
and XTE J1859+226, respectively.  With the addition of these 
2 systems, 14 out of the 15 transients with confirmed black 
holes (via compact object mass measurements) now have 
measured quiescent luminosities or sensitive upper limits.
The only exception is GRS 1915+105, which has not been 
in quiescence since its discovery in 1992.  The luminosities
for V4641~Sgr and XTE J1859+226 are consistent with the
median luminosity of $2\times 10^{31}$ erg~s$^{-1}$ for
the systems with previous detections.  Our analysis suggests 
that the quiescent X-ray spectrum of V4641~Sgr is harder than 
for the other systems in this group, but, due to the low 
statistical quality of the spectrum, it is not clear if 
V4641~Sgr is intrinsically hard or if the column density is 
higher than the interstellar value.  Focusing on V4641~Sgr, we 
compare our results to theoretical models for X-ray emission 
from black holes in quiescence.  Also, we obtain precise X-ray
positions for V4641~Sgr and XTE J1859+226 via cross-correlation
of the X-ray sources detected near our targets with IR sources 
in the 2 Micron All-Sky Survey catalog.

\end{abstract}


\section{Introduction}

Previous measurements of black hole (BH) X-ray transients in 
quiescence show that the BH systems have X-ray luminosities ranging 
from $10^{30}$ to $10^{33}$ erg~s$^{-1}$ \citep{garcia01,hameury03}.
As in outburst, there is evidence for the presence of an accretion 
disk in quiescence \citep{ob97,mcclintock03}, but the accretion rate 
onto the BH itself is unclear as is the nature of the accretion flow.  
Although there is currently no consensus on the origin of the 
quiescent X-ray emission, three emission sites are considered to be 
viable for at least some systems:  The accretion disk; the putative 
outflow or jet; and the secondary star.

One picture for the quiescent accretion flow is that of an 
inner quasi-spherical and optically thin region surrounded by
a optically thick disk as in the advection-dominated accretion 
flow (ADAF) model \citep{nmy96}.  This model explains the 
faintness of quiescent BHs as being due to advection of 
radiation across the BH event horizon \citep{garcia01}.  
However, it has been determined that ADAFs are convectively 
unstable for certain viscosity ranges, leading to 
convection-dominated accretion flows (CDAF, Quataert \& 
Gruzinov 2000\nocite{qg00}).  Such a model can explain the 
low BH luminosities without the need for advection across an 
event horizon.  In addition, BH advection may not be necessary 
if outflows or jets transport the accretion energy out of the 
system \citep{bb99,gbg99,fgj03}.  There is evidence for the 
presence of compact radio jets from BH systems at low X-ray 
luminosities \citep{fender01,corbel03}, and these may persist 
in quiescence.  Such jets may possibly contribute to the X-ray 
emission \citep{mff01}.  Finally, \cite{br00} suggested the 
secondary as an X-ray emission site, but we note that this 
possibility has been ruled out for most systems \citep{kong02a}.

In this work, we describe results of observations of the BH 
transients V4641~Sgr and XTE J1859+226 in quiescence made using the 
{\em Chandra X-ray Observatory} \citep{weisskopf02}.  These systems 
both had major X-ray outbursts in 1999 \citep{wood99,slm99}.  While 
the XTE J1859+226 outburst was typical of BH transients, the behavior 
of V4641~Sgr was unusual with a long period of low level activity 
punctuated by a 12 Crab flare lasting 1-2 days.  From optical 
observations covering the 67.6 hr V4641~Sgr orbit and the (likely) 
9.16 hr XTE J1859+226 orbit, the compact object masses are above 
3\Msun, indicating that these systems contain BHs \citep{orosz01,fc01}.  
At the time of the outburst, V4641~Sgr exhibited a one-sided, rapidly 
evolving, relativistic radio jet \citep{hjellming00,orosz01}, and 
there is evidence that XTE J1859+226 also produced a radio jet 
\citep{brocksopp02}.  The main motivations of our {\em Chandra} 
observations are to study the BHs in quiescence and to search for 
large-scale X-ray jets like those detected for XTE J1550--564 
\citep{corbel02} and 4U 1755--338 \citep{aw03}.  Here, we report 
on the detection of the BHs in quiescence, and we will report on 
the X-ray jet search and the consequences of the non-detection of 
those jets in a future paper.

\section{Observations}

We obtained 2 {\em Chandra} observations of V4641~Sgr in 2002:  
A 4.3 ks observation on UT August 5 (obs. 1a); and a 25.3 ks 
observation on UT October 21 (obs. 1b).  For both observations, 
we used the Advanced CCD Imaging Spectrometer (ACIS) in imaging 
mode, and the target was placed on one of the back-illuminated 
ACIS chips (S3).  Observation 1a was carried out under Director's 
Discretionary Time and was prompted by X-ray, optical, and 
radio activity from the source during 2002 May and June 
\citep{ms02,uemura02,rdm02}.  Considering the possibility that 
the source might still be active in X-rays during observation 1a, 
we used a 1/4 CCD subarray to mitigate photon pileup effects.  
We found that the source was not bright \citep{tomsick_atel02}, 
and we used the full CCD for observation 1b.  We also obtained 
a 24.8 ks {\em Chandra} observation of XTE J1859+226 
on 2003 February 5 (obs. 2) using ACIS in imaging mode.  
The last known X-ray activity from XTE J1859+226 was in 2000 
\citep{th00}.  For all 3 observations, the background levels 
remain low throughout, allowing us to use the full exposure time.
VLA (Very Large Array) radio observations occurred within
1 week of observations 1a and 1b and within 2 weeks of 
observation 2.  V4641~Sgr and XTE J1859+226 showed no
radio emission with rms noise levels near 0.1 mJy/beam
at 4 and 6 cm.

\section{Source Detections and 2MASS/{\em Chandra} Cross-Correlation}

\begin{figure}
\centerline{\includegraphics[width=0.40\textwidth]{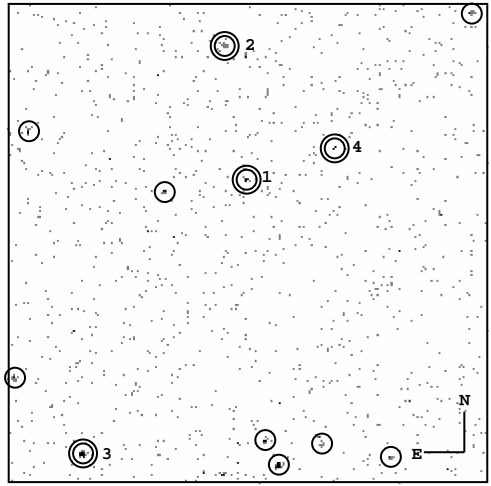}}
\caption{0.3-8 keV {\em Chandra} image of the V4641~Sgr field.  The 
exposure time is 29.5 ks, and the image size is $1.968$ arcmin$^{2}$.  
The N and E lines are $10^{\prime\prime}$.  The 12 detected sources 
are circled, and sources 1-4 (marked with 2 circles) have 2MASS 
identifications.  Source 1 is V4641~Sgr.  
\label{fig:v4641}}
\end{figure}

We produced 0.3-8 keV ACIS images using the ``level 2'' event 
lists from the standard data processing.  For the analysis described 
below, we used the CIAO ({\em Chandra Interactive Analysis of Observations}) 
v2.3 software and CALDB (Calibration Data Base) v2.21.  To obtain the 
maximum sensitivity for source detection in the V4641~Sgr field, we 
combined the data for observations 1a and 1b.  We restricted our search 
for sources to a 2 arcmin$^{2}$ region containing the position of 
V4641~Sgr.  This region, which is shown in Figure~\ref{fig:v4641}, is 
fully covered by both observations giving a spatially uniform exposure 
time of 29.5 ks.  We used the CIAO routine ``wavdetect'' with a detection 
threshold of $10^{-6}$ \citep{freeman02} to search for sources in the 
combined image, leading to the detection of 12 sources with between 
6 and 36 counts per source.  A source with 9 counts (labeled 1 in 
Figure~\ref{fig:v4641}) is detected at a position consistent with the 
V4641~Sgr radio position.

As the V4641~Sgr radio position is known to 
$0^{\prime\prime}.1$ \citep{hjellming00}, it is worthwhile
to obtain the best possible X-ray position in order to test 
whether {\em Chandra} source 1 is, in fact, V4641~Sgr.  
To register the image, we cross-correlated the {\em Chandra} 
source positions with the 2 Micron All-Sky Survey (2MASS) IR 
sources in the field.  We determined that 4 of the 12 
{\em Chandra} sources have 2MASS sources within the 
{\em Chandra} pointing uncertainty of $0^{\prime\prime}.6$.  
For these 4 sources, the angular separations between the 2MASS 
and {\em Chandra} positions range from $0^{\prime\prime}.03$ 
to $0^{\prime\prime}.25$.  Given the surface density of 2MASS 
sources ($7\times 10^{-3}$ sources arcsec$^{-2}$ down to 
K$_{s}\sim 14$), there is a 0.14\% probability that a match 
with the largest separation is spurious.  As source 1 (the 
V4641~Sgr candidate) is identified as one of the 2MASS 
sources (2MASS 18192163--2524258 with IR magnitudes 
J = $12.532\pm 0.029$, H = $12.364\pm 0.027$, and 
K$_{s}$ = $12.270\pm 0.030$), we used the other 3 sources to 
register the image.  For these 3 sources, the average 2MASS
to {\em Chandra} differences in R.A. and Decl. are 
$-0.05\pm 0.13$ and $0.10\pm 0.13$ arcseconds, respectively, 
where the uncertainties account for the {\em Chandra} 
statistical position error as well as the $0^{\prime\prime}.2$ 
position uncertainties for 2MASS\footnote{See 
http://www.ipac.caltech.edu/2mass/releases/second/doc/}.  
We performed the indicated shifts ($0^{\prime\prime}.05$ East 
and $0^{\prime\prime}.1$ South) to complete the registration of 
the {\em Chandra} image to the 2MASS positions.  The position of 
source 1 is R.A. = 18h 19m 21s.641, Decl. = --25$^{\circ}$ 
24$^{\prime}$ 25$^{\prime\prime}.87$ (equinox 2000.0, systematic 
uncertainty = $0^{\prime\prime}.13$, statistical uncertainty = 
$0^{\prime\prime}.13$).  The best {\em Chandra} position 
is $0^{\prime\prime}.12$ from the 2MASS position and 
$0^{\prime\prime}.06$ from the radio position obtained from a 
VLA image obtained on 1999 September 17, for which 
\cite{hjellming00} interpreted the emission as coming from the 
black hole rather than the extended jet.  After registering the 
image, we find that all 9 ACIS counts are contained within 
$1^{\prime\prime}$ of the radio position, while the expected 
number of background counts in a $1^{\prime\prime}$ circle is 
0.28.  Thus, the Poisson probability that the detection is 
spurious is $2\times 10^{-11}$.

\begin{figure}
\centerline{\includegraphics[width=0.37\textwidth]{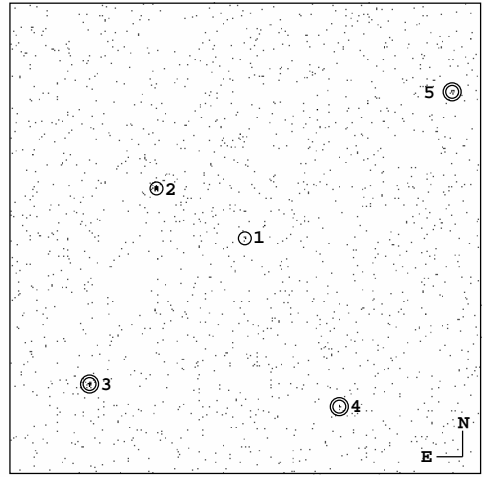}}
\caption{0.3-8 keV {\em Chandra} image of the XTE J1859+226 field.  
The exposure time is 24.8 ks, and the image size is 3 arcmin$^{2}$.  
The N and E lines are $10^{\prime\prime}$.  The 5 detected sources
are circled, and and the 3 sources marked with 2 circles have 2MASS 
identifications.  Source 1 is XTE J1859+226.\label{fig:1859}}
\end{figure}

We carried out a similar analysis for XTE J1859+226.  We 
searched for {\em Chandra} sources in a 3 arcmin$^{2}$ 
region centered on the XTE J1859+226 radio position \citep{ph99}
using wavdetect.  As shown in Figure~\ref{fig:1859}, we 
detected 5 {\em Chandra} sources with between 6 (source 1) and 
113 (source 2) counts.  Source 1 is consistent with the 
XTE J1859+226 radio position.  We cross-correlated the 
{\em Chandra} positions with the 2MASS positions and found 
that sources 3-5 have likely IR counterparts.  For these 3 
sources, the maximum separation between the {\em Chandra} and 
2MASS positions is $0^{\prime\prime}.33$, and there is a 
0.07\% probability that a match with this separation is spurious.  
Using the 2MASS identifications, we determined the position 
shifts required to register the {\em Chandra} image, and these
are $0.09\pm 0.12$ and $0.25\pm 0.14$ arcseconds
in R.A. and Decl., respectively.  After shifting the 
{\em Chandra} image by $0^{\prime\prime}.09$ West and 
$0^{\prime\prime}.25$ South, the position of source 1
is R.A. = 18h 58m 41s.485, Decl. = +22$^{\circ}$ 39$^{\prime}$ 
29$^{\prime\prime}.88$ (equinox 2000.0, systematic uncertainty 
is $0^{\prime\prime}.14$, statistical uncertainty is 
$0^{\prime\prime}.10$ in R.A. and $0^{\prime\prime}.20$ in Decl.), 
which is consistent with the $0^{\prime\prime}.3$ VLA radio
position \citep{ph99}.  All 6 ACIS counts are contained within 
$1^{\prime\prime}$ of the radio position, while the expected 
number of background counts in a $1^{\prime\prime}$ circle 
is 0.18.  Thus, the Poisson probability that the detection is 
spurious is $4\times 10^{-8}$.

\section{Quiescent Black Hole Properties}

Although the number of counts detected for both V4641~Sgr
and XTE J1859+226 is small, we extracted energy spectra
(with 9 and 6 counts, respectively).  For V4641~Sgr, the 
photon energies range from 0.87 keV to 7.0 keV, and the 
mean photon energy is 3.5 keV.  For XTE J1859+226, the 
photon energies range from 0.75 keV to 2.7 keV, and the 
mean photon energy is 1.5 keV.  We fitted the energy 
spectra with a power-law model with interstellar absorption.  
We used this model because this spectral shape is typically 
seen for quiescent BH systems \citep{garcia01,kong02a}, 
but we emphasize that we cannot rule out other spectral 
shapes.  We fitted the spectra using Cash statistics 
\citep{cash79}, and we fixed the column density ($N_{\rm H}$) 
to the H~I values from \cite{dl90}: 
$2.3\times 10^{21}$ cm$^{-2}$ and $2.2\times 10^{21}$ cm$^{-2}$ 
for V4641~Sgr and XTE J1859+226, respectively.  We determined 
the errors on the parameters by producing and fitting 10,000
simulated spectra.  We used the best fit parameters from the 
fits to the actual data as input to the simulations.  The 
power-law photon indices are $\Gamma = 0.2^{+0.9}_{-1.0}$
for V4641~Sgr and $\Gamma = 2.4^{+1.5}_{-1.3}$ for
XTE J1859+226 (90\% confidence errors).  

The energy spectra of quiescent BHs are typically 
well-described by an absorbed power-law with a photon index 
of $\Gamma = 1.3$-2.3 \citep{kong02a,hameury03,mcclintock03}.
Although a (much) better V4641~Sgr spectrum should be 
obtained to confirm its spectral shape, the best estimate of 
$\Gamma = 0.2$ is well outside of the typical range, and even 
at the upper limit of $\Gamma = 1.1$, V4641~Sgr would be the 
hardest of the quiescent BHs.  However, the uncertain column 
density is an important source of systematic error, and we 
re-fitted the V4641~Sgr spectrum leaving $N_{\rm H}$ as a 
free parameter.  Our simulations indicate 90\% confidence 
upper limits on $\Gamma$ and $N_{\rm H}$ of 2.5 and 
$3.3\times 10^{22}$ cm$^{-2}$, respectively.  Thus, we
conclude that either the V4641~Sgr is intrinsically hard
or that $N_{\rm H}$ is higher than the interstellar value.

We used the spectral fits to obtain flux and luminosity 
measurements for the sources.  For V4641~Sgr, the absorbed 
0.3-8 keV flux is $(6.5^{+5.7}_{-4.0})\times 10^{-15}$
erg~cm$^{-2}$~s$^{-1}$ (90\% confidence errors), and the 
unabsorbed luminosity in the same energy band is 
$(4.0^{+3.3}_{-2.4})\times 10^{31}$ ($d$/7 kpc)$^{2}$ 
erg~s$^{-1}$, where $d$ is the source distance.  Optical 
observations of V4641~Sgr in quiescence give a distance 
of $7.3\pm 1.2$ kpc (Orosz et al., in prep.).  For 
XTE J1859+226, the absorbed 0.3-8 keV flux is 
$(1.5^{+0.2}_{-0.7})\times 10^{-15}$ erg~cm$^{-2}$~s$^{-1}$, 
and the luminosity is $(4.2^{+4.8}_{-2.2})\times 10^{31}$ 
($d$/11 kpc)$^{2}$ erg~s$^{-1}$.  The distance estimate of 
11 kpc comes from the X-ray and optical properties of the
source \citep{zurita02}.  

\section{Discussion}

\begin{figure}
\centerline{\includegraphics[width=0.5\textwidth]{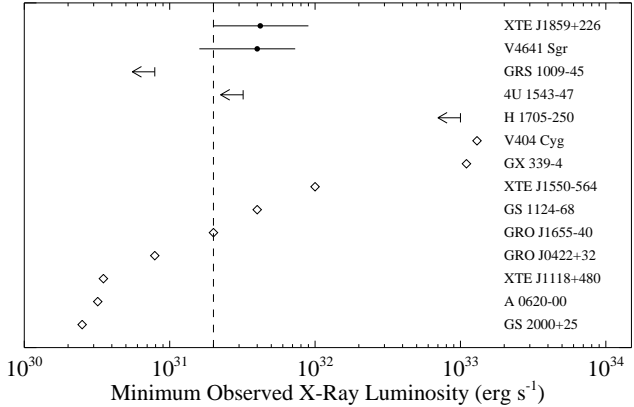}}
\caption{Quiescent black hole X-ray luminosities.  The points with 
error bars mark the 0.3-8 keV luminosities of XTE J1859+226 and 
V4641~Sgr from our {\em Chandra} measurements, assuming distances 
of 11 kpc and 7 kpc for the two sources, respectively.  The diamonds 
and upper limits are the minimum X-ray luminosities reported for the 
other 12 BH systems for which quiescent measurements have been possible
\citep{garcia01,hameury03,sutaria02,mcclintock03,kong02b,kaaret03,corbel03}.
The dashed line is the median luminosity for the previous
detections.\label{fig:luminosity}}
\end{figure}

As confirmed black hole systems (based on compact object 
mass measurements), V4641~Sgr and XTE J1859+226 represent 
important additions to the group of BHs that have 
previously been detected in X-rays in quiescence.  Of the 
15 confirmed \citep{mr03} transient BH systems, 14 now 
have quiescent X-ray detections or sensitive upper limits, 
and only GRS 1915+105, which has been in outburst since its 
discovery in 1992, does not.  The X-ray luminosity ($L_{x}$) 
measurements are summarized in Figure~\ref{fig:luminosity}.  
We include a new low X-ray flux measurement for XTE 
J1550--564 \citep{kaaret03}, that has not been previously 
considered in the context of studies of quiescent BHs, and 
we assume a distance of 5.3 kpc \citep{orosz02} to convert 
from flux to luminosity.  We also include the lowest flux 
measurements for the recurrent transient GX 339--4 
\citep{kong02b,corbel03}, and we assume a distance of 4 kpc.  
The median luminosity is $2\times 10^{31}$ erg~s$^{-1}$ for 
the 9 systems with previous detections, and the measured 
luminosities for V4641~Sgr and XTE J1859+226 are consistent 
with the median.

For quiescent BHs, the X-ray emission is thought to originate 
from the accretion flow, a jet, or the secondary star, and, 
here, we discuss how our results for V4641~Sgr may constrain 
theoretical models for X-ray production.  The V4641~Sgr source 
distance (7 kpc) along with the proper motion of the one-sided
radio jet seen in 1999 ($0^{\prime\prime}.36$ per day) indicate 
that the angle between the jet axis and our line of sight is 
$<$8$^{\circ}$ \citep{orosz01}.  Thus, if the X-ray emission 
originates in a fast-moving jet, one expects V4641~Sgr to be
brighter due to relativistic beaming.  However, the V4641~Sgr
X-ray luminosity (assuming isotropic emission) is similar to 
the luminosities of the other BH systems, for which the jet 
axes are either not known or are known to be relatively far 
from our line of sight, indicating that the X-ray emission is 
not highly beamed and limiting the velocity of a putative 
X-ray emitting jet.  For a continuous synchrotron X-ray jet 
\citep{mff01}, it is unlikely that the bulk-motion Lorentz 
factor could be higher than $\sim$1.5 as this would cause 
the source to be brighter than an unbeamed source by a factor 
of $\sim$10 for a jet that is $8^{\circ}$ from our line of 
sight \citep{mr99}.  This calculation assumes a spectrum with 
a photon index of 1.5, which is in the range of values that 
can be produced within the \cite{mff01} model.  Assuming a 
photon index of 0.2 (our best estimate for V4641~Sgr) would 
lead to a somewhat higher limit on the Lorentz factor, but it 
is unclear whether such a hard spectrum could have a synchrotron 
origin.  Also, we note that the Lorentz factor constraint is 
similar for models where X-rays are produced in jets via inverse 
Comptonization \citep{gak02}.  The V4641~Sgr constraint on the 
Lorentz factor ($\lesssim$1.5) is consistent with recent results 
for BH systems in the canonical low-hard state 
\citep{gfp03,maccarone03}.

For accretion disk models such as the ADAF model, the quiescent 
X-ray luminosity depends mainly on mass accretion rate rather 
than on system orientation.  Thus, the luminosities of V4641~Sgr 
and XTE J1859+226 would be expected to be similar to the other 
BH systems, as observed.  If the hard V4641~Sgr X-ray spectrum 
is confirmed and is intrinsic to the source, it would suggest 
that another parameter besides accretion rate is important.  
Hard spectra can be produced by ADAFs \citep{mcclintock03} 
and CDAFs \citep{qg00}, but it is unclear why one system at 
close to the median luminosity would be intrinsically much 
harder than the others.

The spectral type of the V4641~Sgr secondary is B9~III
\citep{orosz01}, making it the most luminous of the BH
transients.  In quiescence, the ratio of the X-ray to 
bolometric luminosity ($L_{x}/L_{bol}$) for V4641~Sgr is 
$\sim$$3\times 10^{-5}$ based on the X-ray luminosity reported 
here and the optical luminosity of the secondary (Orosz et al., 
in prep.).  The V4641~Sgr ratio is higher than the average 
value but comparable to the highest values measured by {\em ROSAT} 
for late B-type stars \citep{berghoefer97}.  While it is unlikely 
that the hard X-ray flux from V4641~Sgr comes from the secondary, 
we cannot rule out the possibility that the secondary makes a 
some contribution to the soft X-ray flux below $\sim$2~keV.  
The early type secondary could also be important if it has a
strong wind.  The wind could collide with accretion disk material, 
leading to additional X-ray production, or it could cause extra 
X-ray absorption, explaining the hard V4641~Sgr spectrum.

\acknowledgments

We would like to thank H. Tananbaum for granting Director's
Discretionary Time and the referee for very useful comments.
The Two Micron All Sky Survey is a joint project of the Univ. 
of Massachusetts and IPAC/Caltech, funded by NASA and NSF.  
The National Radio Astronomy Observatory is a facility of the 
NSF operated under cooperative agreement by Associated 
Universities, Inc.  JAT acknowledges partial support from 
{\em Chandra} award number GO3-4040X.  PK acknowledges partial 
support from NASA grant NAG5-7405 and {\em Chandra} award 
number GO3-4043X.  JMM thanks the NSF.


\end{document}